\documentstyle[cjaa209,twoside,epsf]{article}
\input{psfig.sty}

\begin{document}

\newcommand{\kms}{km~s$^{-1}$}
\newcommand{\Msun}{$M_\odot$}
\newcommand{\Msunn}{$M_\odot$ }
\def\simlt{\lower.5ex\hbox{$\; \buildrel < \over \sim \;$}}
\def\simgt{\lower.5ex\hbox{$\; \buildrel > \over \sim \;$}}
\def\arcdeg{\hbox{$^\circ$}}
\def\arcmin{\hbox{$^\prime$}}
\def\arcsec{\hbox{$^{\prime\prime}$}}

\title{{\textsl{UBVI}} Surface Photometry of the Spiral Galaxy 
      NGC 300 in the Sculptor Group}

\author{Sang Chul Kim$^1$\mailto{}, Hwankyung Sung$^2$, 
   Hong Soo Park$^3$, and Eon-Chang Sung$^1$}

\inst{
$^1$Korea Astronomy Observatory, Taejon 305-348, Republic of Korea \\
$^2$Department of Astronomy and Space Science, Sejong
     University, 98 Gunja-dong, Gwangjin-gu, Seoul 143-747, Republic of Korea \\
$^3$Astronomy Program, SEES, Seoul National University,
     Seoul 151-742, Republic of Korea}

\email{sckim@kao.re.kr}

\markboth{S. C. Kim et al.}
{Surface Photometry of NGC 300}

\pagestyle{myheadings}

\date{Received~~2004~~~~~~~~~~~~~~~ ; accepted~~2004~~~~~~~~~~~~~~ }

\baselineskip=18pt

\begin{abstract}
\baselineskip=18pt
We present {\it UBVI} surface photometry for $20.'5 \times 20.'5$ area
of a late-type spiral galaxy NGC 300. 
In order to understand the morphological properties and luminosity distribution
characteristics of NGC 300,
we have derived isophotal maps, surface brightness profiles, 
ellipticity profiles, position angle profiles, and color profiles.
By merging the $I$-band data of our surface brightness measurements 
with those of B\"oker et al. (2002) based on {\it Hubble Space Telescope}
observations, we have made combined $I$-band surface brightness profiles for
the region of 0.\arcsec02 $< r <$ 500\arcsec~ and 
decomposed the profiles into three components: a nucleus, a bulge,
and an exponential disk.
\end{abstract}

\section{Introduction} 

NGC 300 (=ESO 295-G020, IRAS 00525-3757, PGC 3238) is a late-type (SA(s)d, 
de Vaucouleurs et al. 1991) spiral
galaxy in the nearest galaxy group, 
the Sculptor group, which contains five major spiral galaxies (NGC 55, 
NGC 247, NGC 253, NGC 300, and NGC 7793) and $\sim$20 dwarf galaxies
(C{\^o}t{\'e} et al. 1997, Whiting 1999, Karachentsev et al. 2003; 
cf. van den Bergh 1999). 
NGC 300 is a rather bright ($M_B = -18.6$) and nearly face-on galaxy,
similar to M33 in the northern hemisphere (Blair \& Long 1997).
Some basic information of this galaxy is summarized in Table 1,
which is supplemented to the table 1 of Kim, Sung, \& Lee (2002, hereafter Paper I).

\begin{table*}
\caption{Basic Information of NGC 300}
\begin{center}
{\scriptsize
\begin{tabular}{llc}
  \hline
  Parameter & Information & Reference \\
  \hline
$\alpha_{J2000}$, $\delta_{J2000}$ &
        0$^h$ 54$^m$ 53.$^s$48, $-37$\arcdeg~ $41'$ $03.''8$ & NASA/IPAC Extragalactic Database \\
$l, b$             & 299.\arcdeg21, $-79.\arcdeg42$ & NASA/IPAC Extragalactic Database \\
Type               & SA(s)d                        & de Vaucouleurs et al. 1991 \\
{\sc{H~i}} heliocentric radial velocity, $v_\odot$ &
                                142 $\pm$ 4 km s$^{-1}$ & de Vaucouleurs et al. 1991 \\
Maximum rotational velocity, $V_{max}$ & 87 km s$^{-1}$ & Carignan 1985 \\
Foreground reddening, $E(B-V)$ & 0.013 mag & Schlegel et al. 1998 \\
Isophotal major diameter at the 25.0 $B$ mag arcsec$^{-2}$ level, D$_{25}$ &
                                21.$'$9 & de Vaucouleurs et al. 1991 \\
Position angle 
                   & 111\arcdeg  & de Vaucouleurs et al. 1991 \\
Minor to major axis ratio at D$_{25}$, $(b/a)_{25}$  &  0.74 & Carignan 1985 \\
Inclination $i$    & $42.\arcdeg3 \pm 3.\arcdeg0$ & Carignan 1985 \\
Distance modulus, (m-M)$_0$ & $26.53 \pm 0.07$ mag & Freedman et al. 2001 \\
                            &                      & (cf. Butler et al. 2004) \\
Distance, $d$ & $2.02\pm 0.07$ Mpc (1$''$ = 9.8 pc) & Freedman et al. 2001 \\
Exponential disk corrected central surface brightness, $\mu_B(0)$ & 22.23 mag arcsec$^{-2}$ &
                                Carignan 1985 \\
Exponential disk corrected central surface brightness, $\mu_I(0)$ & 19.97 mag arcsec$^{-2}$ &
                                This study \\
Scale length, $r_s$ (obtained from $B$-band) & 2.06 kpc            & Carignan 1985 \\
Scale length, $r_s$ (obtained from $I$-band) & 1.47 kpc            & This study \\
Apparent total magnitude, $B_{\rm{T}}$       & $8.72 \pm 0.05$ mag & de Vaucouleurs et al. 1991 \\
Apparent total color index, $(B-V)_{\rm{T}}$ & $0.59 \pm 0.03$ mag & de Vaucouleurs et al. 1991 \\
Apparent total color index, $(U-B)_{\rm{T}}$ & $0.11 \pm 0.03$ mag & de Vaucouleurs et al. 1991 \\
Corrected total magnitude, $B_{\rm{T}}^0$    & 8.53                & de Vaucouleurs et al. 1991 \\
Corrected total color index, $(B-V)^0_{\rm{T}}$ &  0.56            & de Vaucouleurs et al. 1991 \\
Corrected total color index, $(U-B)_{\rm{T}}^0$ &  0.09             & de Vaucouleurs et al. 1991 \\
Absolute total magnitude, $M_{B_{T}}$ & $-18.59$ & Sandage \& Tammann 1981 \\
Metallicity from the ratio of carbon stars to M stars of & & \\
        \hspace*{2cm} spectral type M5 or later, [Fe/H] &
                                $-0.5$ dex & Richer, Pritchet, \& Crabtree 1985 \\
{\sc{H~i}} column density in the direction of NGC 300 & $2.97 \times 10^{20}$ atoms cm$^{-2}$ &
        Read, Ponman, \& Strickland 1997 \\
{\sc{H~i}} flux            & $\leq 670$ Jy km s$^{-1}$ & Carignan 1985 \\
{\sc{H~i}} mass, $M_{H I}$ & $(2.5 \pm 0.1) \times 10^9$ $M_\odot$ &
                                Rogstad, Crutcher, \& Chu 1979 \\
\hline
\end{tabular}
} 
\end{center}
\end{table*}

Although there have been several studies on various objects in NGC 300
(e.g. Paper I and references therein),
only a few studies exist on surface photometry study on this galaxy.
De Vaucouleurs \& Page (1962) performed surface photometry study of NGC 300
using photographic plates obtained at Mount Stromlo 30- and 74-inch reflectors.
Carignan (1985) has performed surface photometry of the three Sculptor group 
galaxies: NGC 7793, NGC 247, and NGC 300.
He has used $B_J$-band photographic plates taken by using the 1.2m Schmidt telescope
at Siding Spring Observatory.
Recently, B\"oker et al. (2002, 2004) have observed 77 nearby late-type spiral galaxies
including NGC 300 using the {\it Hubble Space Telescope (HST)}/WFPC2.
They have used the $I$-band images of the galaxies to study the nuclear star 
clusters and identified distinct, compact, and dominant sources at or very close
to the photocenter of the 59 galaxies of their sample, which include NGC 300.
They have presented the $I$-band surface brightness profiles of NGC 300 inside
of $r \sim 15$\arcsec~ and some photometric parameters of the nuclear star cluster
in NGC 300.

  There have not been any surface photometry study on NGC 300 with multi-band 
filters and CCD images.  
In this paper, we present surface photometry based on $UBVI$ CCD images for 
the $20.'5 \times 20.'5$ areas of NGC 300.  
We have derived surface brightness distributions, color distributions, and structural
parameters for $r < 500$\arcsec~ of NGC 300.
This paper is composed as follows: 
Section 2 describes the photometric observations and the data reduction process.
The isophotal maps of NGC 300 are presented in Section 3.
In Section 4 ellipse fitting results are presented, 
and color profiles are presented in Section 5.
In Section 6 we compare the surface photometry results with those of previous 
studies and present the surface brightness profile decomposition results.
Finally, a brief summary is given in Section 7.

\section{Observations and Data Reduction} 
$UBVI$ CCD photometry was performed on 1997 November 23 at
Siding Spring Observatory with the 40 inch (1m) telescope (f/8)
and a thinned SITe 2048 $\times$ 2048 CCD (24$\mu$m pixels).  
The scale was 0.$''$602 pixel$^{-1}$, giving 20.$'$5 on a side. 
Exposure times used in the observations were 
1200 s in $U$, 600 s in $B$, 300 s in $V$, and 120 s in $I$.
The night was photometric and the seeings were
$1.''5, 1.''6, 1.''4,$ and $1.''4$ in the $U, B, V,$ and $I$ images,
respectively.

All the preprocessing, such as overscan correction, bias subtraction and flat
fielding were done using the IRAF\footnote{IRAF is distributed by the National
Optical Astronomy Observatories, which are operated by the Association of
Universities for Research in Astronomy, Inc., under cooperative agreement 
with the National Science Foundation.}/{\small CCDRED} package.
Surface photometry of NGC 300 from the CCD images have been performed by
using the ellipse fitting task IRAF/{\small SPIRAL} (Surface 
Photometry Interactive Reduction and Analysis Library),
which is an ellipse fitting package developed at Kiso observatory 
for galaxy surface photometry (Ichikawa et al. 1987) and
is essentially the same as that described by Kent (1983).
  Smoothing of the images have been done using variable-width Gaussian beam 
of IRAF/{\small SPIRAL.SMOOTH} task 
to improve the signal-to-noise ratio in the outer regions of the galaxy.

  The instrumental magnitudes were transformed to the standard Johnson-Cousins
$UBVI$ system using the observation data of equatorial standards of 
South African Astronomical Observatory (SAAO) (Menzies et al. 1991) and of
blue and red standards by Kilkenny et al. (1998), both of which were observed
on the same night.
The atmospheric extinction coefficients were determined from the photometry of 
standard stars and the transformation coefficients adopted by Sung \& Bessell
(2000) were used. 
The transformation equations we derived from the photometry of the standard stars are
$u = U + 4.987  + 0.545  \times X_u + u3    \times (U-B)$,
$b = B + 2.542  + 0.289  \times X_b +0.101  \times (B-V)$,
$v = V + 2.412  + 0.147  \times X_v -0.072  \times (B-V)$, and
$i = I + 2.869  + 0.090  \times X_i -0.028  \times (V-I)$,
where lowercase and uppercase letters represent, respectively, the instrumental system
and standard system (zeropoint$=25.0$).  $X$ represents the air mass and
u3 is $-0.006$ when $(U-B)\ge 0.0$ and $-0.125$ when $(U-B) < 0.0$.
The rms scatter of the standard stars was 
0.012 mag for $U$, 0.008 mag for $B$, 0.011 mag for $V$, and
0.015 mag for $I$, indicating the night was of good photometric quality.

The sky background values were determined from the mode values of the 
south-east edges of the images which are the farthest and clear regions from the galaxy
(the distance from the center of NGC 300 is $\sim 800$\arcsec).
These sky background values were then subtracted from the individual images.
We have calculated the surface brightnesses for these sky values, obtaining
$21.59 \pm 0.22$,     $22.33 \pm 0.08$, 
$21.58 \pm 0.07$, and $19.04 \pm 0.03$ mag arcsec$^{-2}$ for 
$U$, $B$, $V$, and $I$ band, respectively.
The errors quoted here are the estimated standard deviations of 
background sky brightnesses. 
Considering that the lunar phase when our data were obtained is 23 d after
new Moon (or 7 d before the next new Moon) and that
the airmass of our target is in the range of 1.00 -- 1.02,
our sky brightness estimates seem to be reasonably similar to those of the
Kitt Peak National Observatory (Elias 1994; Kim, Lee, \& Geisler 2000)
or those of Crimea/Hawaii sites (Leinert et al. 1995).


\section{Morphological Characteristics of NGC 300} 

  Greyscale maps of $B$ and $V$ CCD images of NGC 300 are shown 
in Fig. 1 (a) and (b), respectively.
The size of the images is $20.'5 \times 20.'5$.
Most of the brightest stars in the images are foreground stars.
The $B$-band image in Fig. 1 (a) shows the spiral features well,
which manifests several lumps of bright stars that are stellar associations
(Pietrzy\'nski et al. 2001; Paper I).
The central region of NGC 300 shown in Fig. 1 (b) reveals 
that the bulge of this SA(s)d type galaxy is quite small compared to 
the size of the spiral arms, which will be discussed in Section 6.2.

\begin{figure}
   \vspace{2mm}
   \begin{center}
   \plottwo{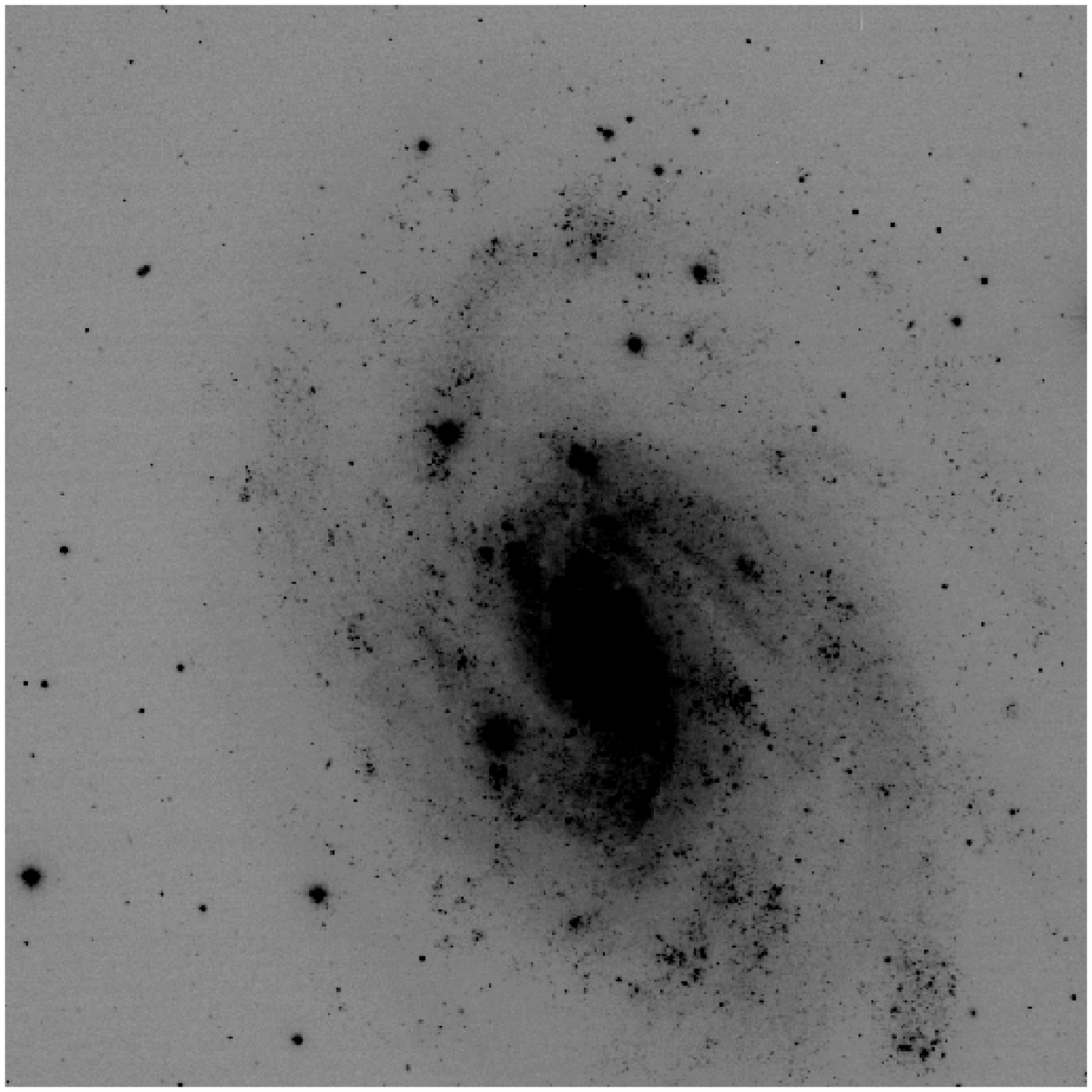}{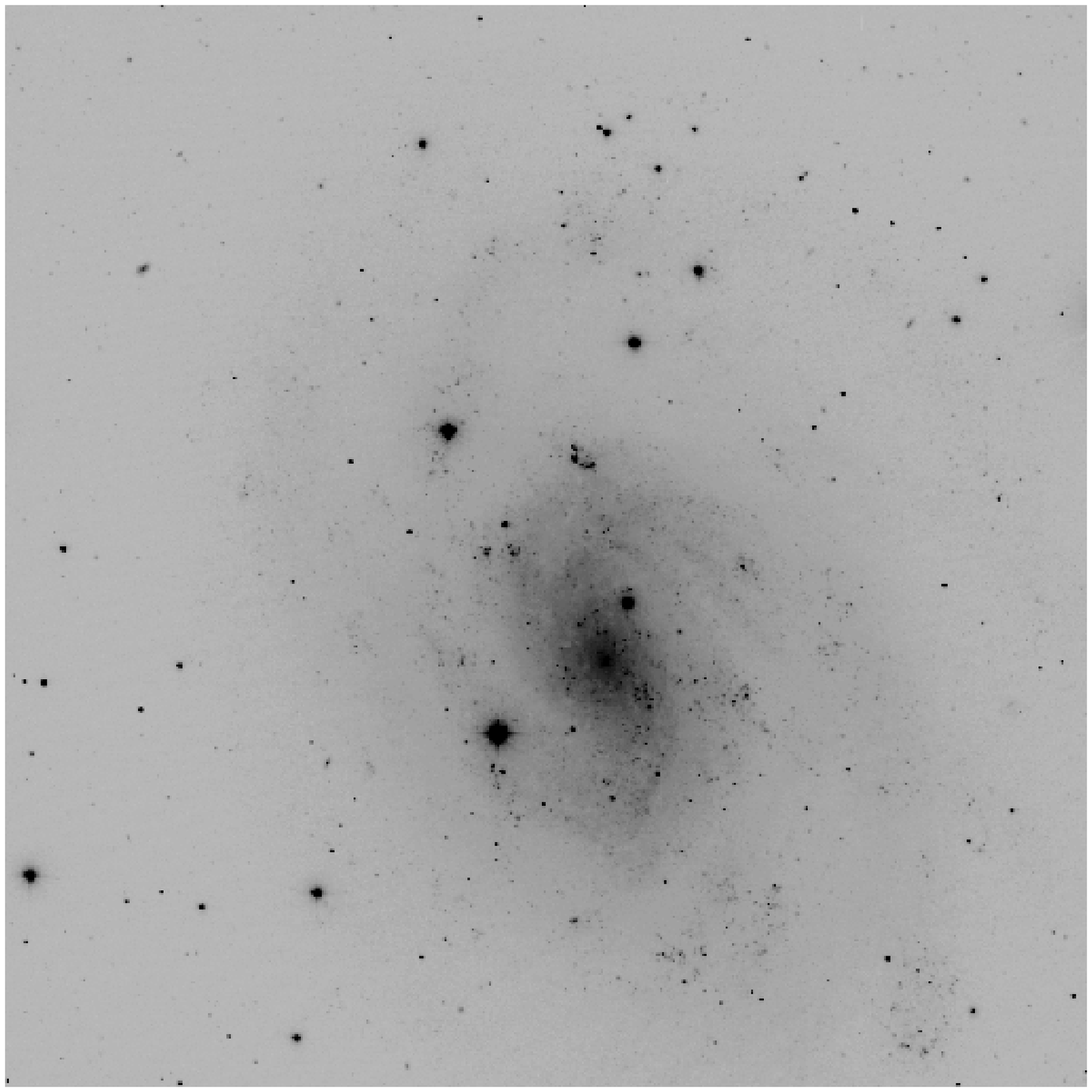}
   \caption{CCD images of NGC 300.
North is to the right of the image and east is at the top.
The scale is 0.$''$602 pixel$^{-1}$ and the size of the field is
$20.'5 \times 20.'5$.
(a: left panel) A greyscale map of the $B$ CCD image of NGC 300.
The spiral arm features are easily seen in this figure.
(b: right panel) A greyscale map of the $V$ CCD image of NGC 300.
This figure shows the central region of the galaxy
with faint features of the spiral arms.
Most of the brightest stars are foreground stars in both of the figures.}
   \end{center}
\end{figure}

  Fig. 2 shows the isophotal maps of NGC 300.
Here the X and Y axes are in pixel units 
and one tick interval corresponds to $60.''2$.
The contour interval is 0.5 mag arcsec$^{-2}$.
North is to the right of the image and east is at the top.
As can be seen in Fig. 2, there is no conspicuous bulge isophotes 
and most isophotes are elongated NW-SE direction excluding the innermost one.
Even though the eastern spiral arm is clearly seen in Fig. 2
(a) and (b), the (north-)western arm is barely delineated thanks to the large 
association of bright stars at around (X=1750, Y=100).
These two major arms are those noted by Sandage \& Bedke (1994).

\begin{figure}
   \begin{center}
   \plotone{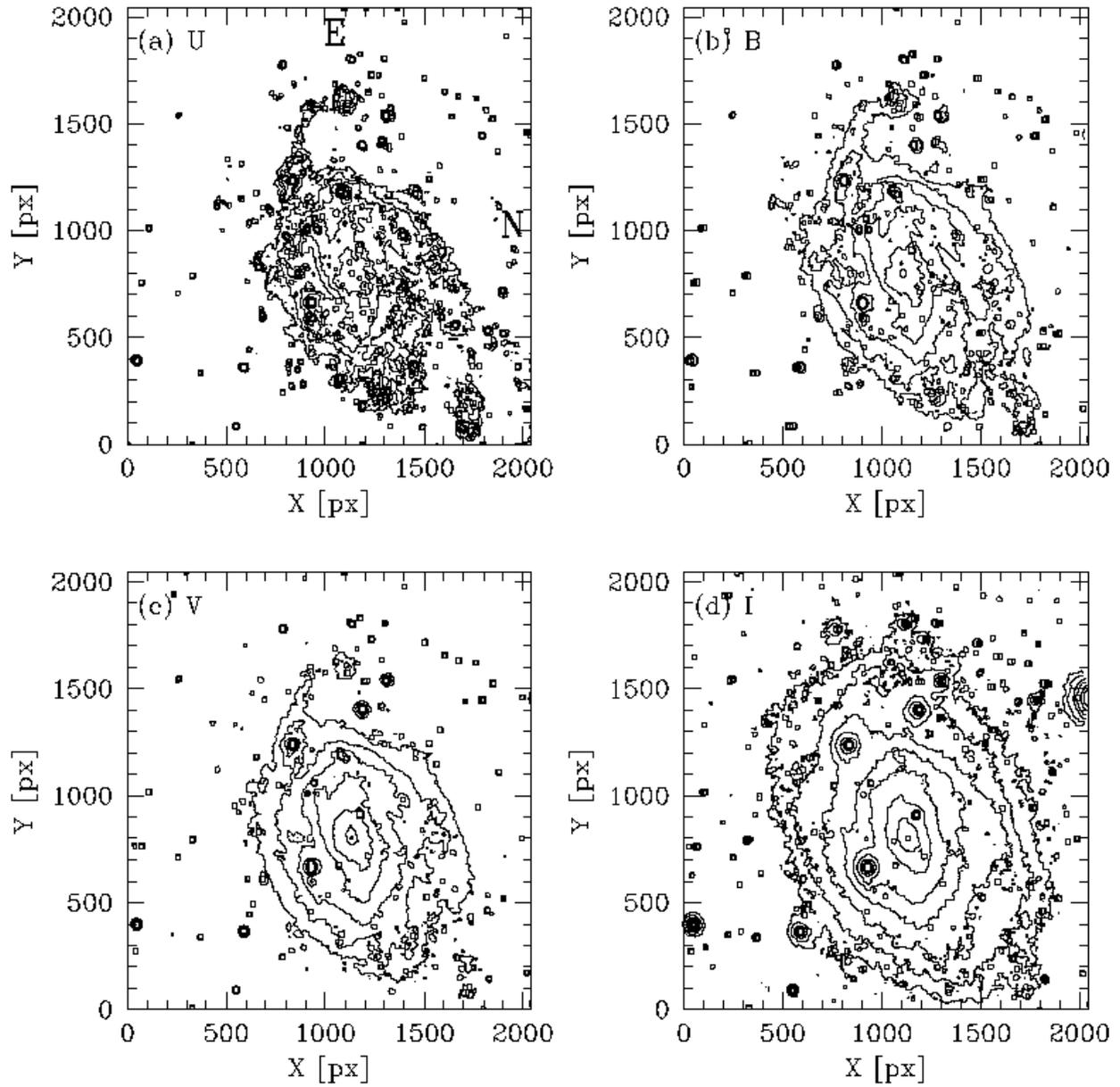}
   \caption{Isophotal contour maps of NGC 300 in (a) $U$-band, (b) $B$-band,
(c) $V$-band, and (d) $I$-band.
North is to the right of the image and east is at the top.
The size of each field is $20.'5 \times 20.'5$
and one tick interval corresponds to $60.''2$.
The contour interval is 0.5 mag arcsec$^{-2}$.}
   \end{center}
\end{figure}

\section{Ellipse Fitting}  
  We have applied the ellipse fitting task IRAF/{\small SPIRAL.PROFS.VPROF} to the NGC 300 
images to obtain surface brightness profiles as well as profiles of ellipticity 
($\epsilon$) and position angle (PA).

Fig. 3 shows $U, B, V$, and $I$ radial surface brightness profiles of 
NGC 300 along the major axis for the region of $r < 500$\arcsec. 
The typical errors of $B$ and $V$-band surface brightnesses
are denoted at the lower and upper panels, respectively.
Since the IRAF/{\small SPIRAL} task does not give the error of 
the surface brightnesses, we have used the 
IRAF/{\small STSDAS.ANALYSIS.ISOPHOTE.ELLIPSE} task
at the given radius using the ellipticity, position angle, and center
coordinates obtained by {\small SPIRAL} task just to obtain the brightness errors.
Unlike the elliptical galaxies which have smooth brightness profiles,
spiral galaxies are patchy and have spiral arm features.
The local patchy features lead to non-systematic error values as in Fig. 3,
though the low photon signals relative to the sky brightness
make the errors get increased at the outer radii.

The surface brightness profiles in $U$ and $B$ are
very similar in their values, which can be deduced
from the apparent total $(U-B)$ color index of NGC 300 of $0.11 \pm 0.03$ mag (de
Vaucouleurs et al. 1991).
While the $U$ surface brightnesses are fainter than those of $B$-band at $r < 200$\arcsec,
they get brighter at $r > 200$\arcsec.

\begin{figure}
   \begin{center}
   \plotone{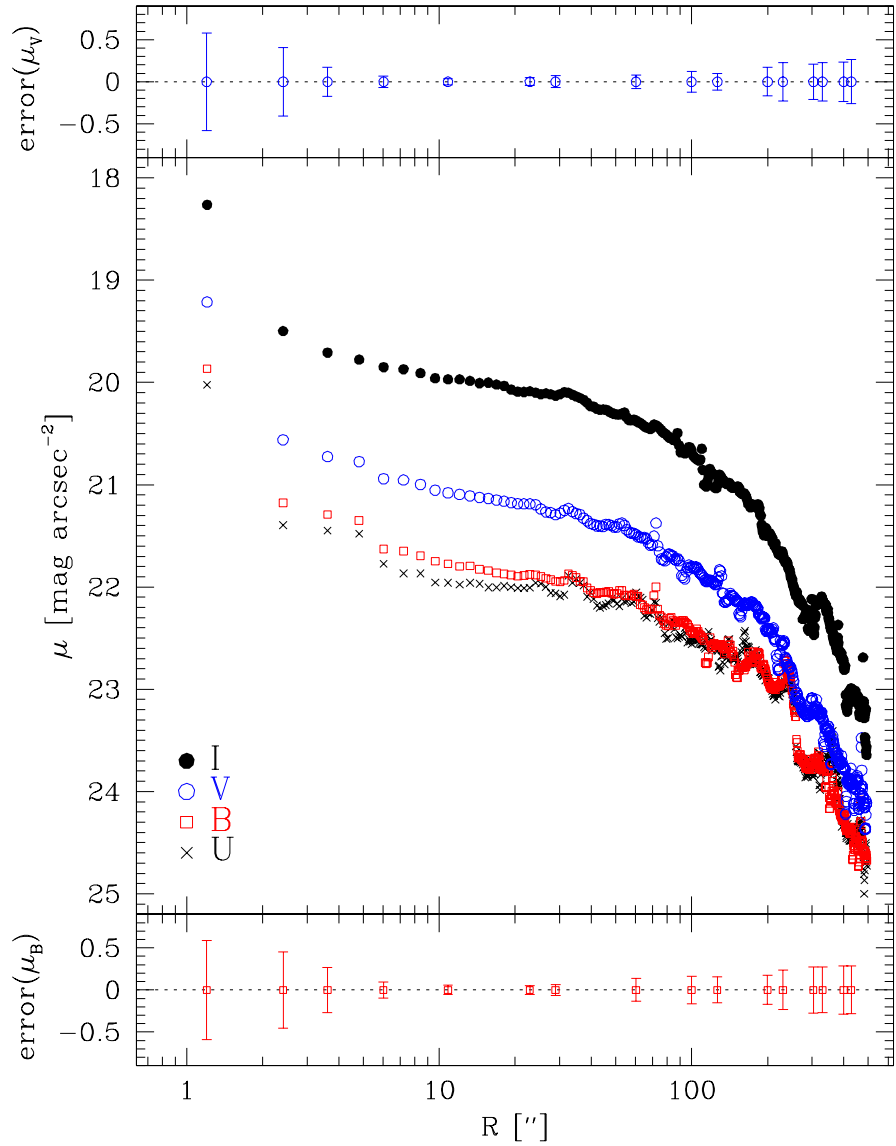}
   \caption{
$U, B, V$, and $I$ major axis surface brightness profiles
of NGC 300.  The typical errors of $B$ and $V$-band surface brightnesses
are denoted at the lower and upper panels, respectively.
{\it The data are available from one of the authors (S.C.K.)}.}
\end{center}
\end{figure}

Fig. 4 shows ellipticity ($\epsilon = 1 - b/a$, 
where $a$ and $b$ are semi-major and semi-minor axis lengths, respectively) 
profiles of NGC 300 in $U, B, V$, and $I$-bands. 
The straight lines are plotted based on eye-fitting and are the same for
all bands.
Ellipticity increases from $\sim$0 (at $r \sim$ 2\arcsec) to $\sim$0.5 (at $r
\sim$ 200\arcsec), and then decreases outward.
The amplitude of ellipticity variation is rather larger in $U$- and $B$-bands
than in the $V$- and $I$-bands, and the absolute values of ellipticities at 
$r > 100$\arcsec~ in $U$- and $B$-bands are also larger than those in 
$V$- and $I$-bands.

\begin{figure}
   \begin{center}
   \plotone{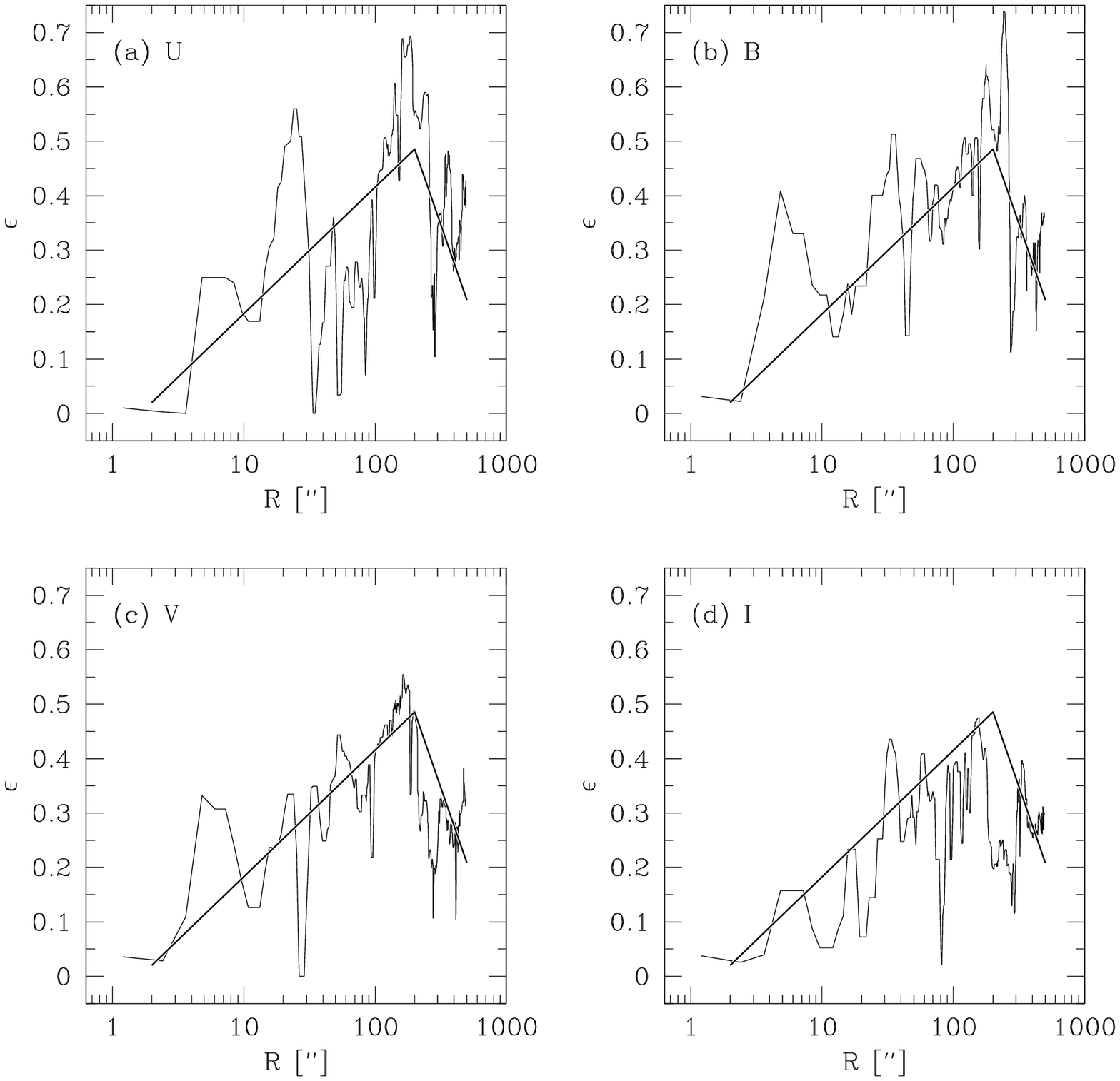}
   \caption{Ellipticity profiles of NGC 300.
The straight lines are based on eye-fitting and are the same for
all bands.}
   \end{center}
\end{figure}

Fig. 5 shows position angle (PA; N through E to the major 
axis) profiles of NGC 300 in $U, B, V$, and $I$-bands. 
The straight lines are plotted based on eye-fitting and are the same for
all bands.
Position angles increase from $\sim$40\arcdeg (at $r \sim$ 1\arcsec~ to 2\arcsec)
to $\sim$110\arcdeg (at $r \sim$ 40\arcsec), and fluctuate around the value of 
$\sim$110\arcdeg~ at regions of $r > 40\arcsec$.
This PA value of $\sim$110\arcdeg~ is in good agreement with those obtained 
by Carignan (1985; 105.\arcdeg6 $\pm$ 1.\arcdeg8) and by de Vaucouleurs et al. 
(1991; 111\arcdeg).
The changes in position angles are mainly due to the spiral structure 
of NGC 300 and there is seen no region of flat PAs (which is an indicator 
of the bar structure; Choi et al. 1998).

   \begin{figure}
   \begin{center}
   \plotone{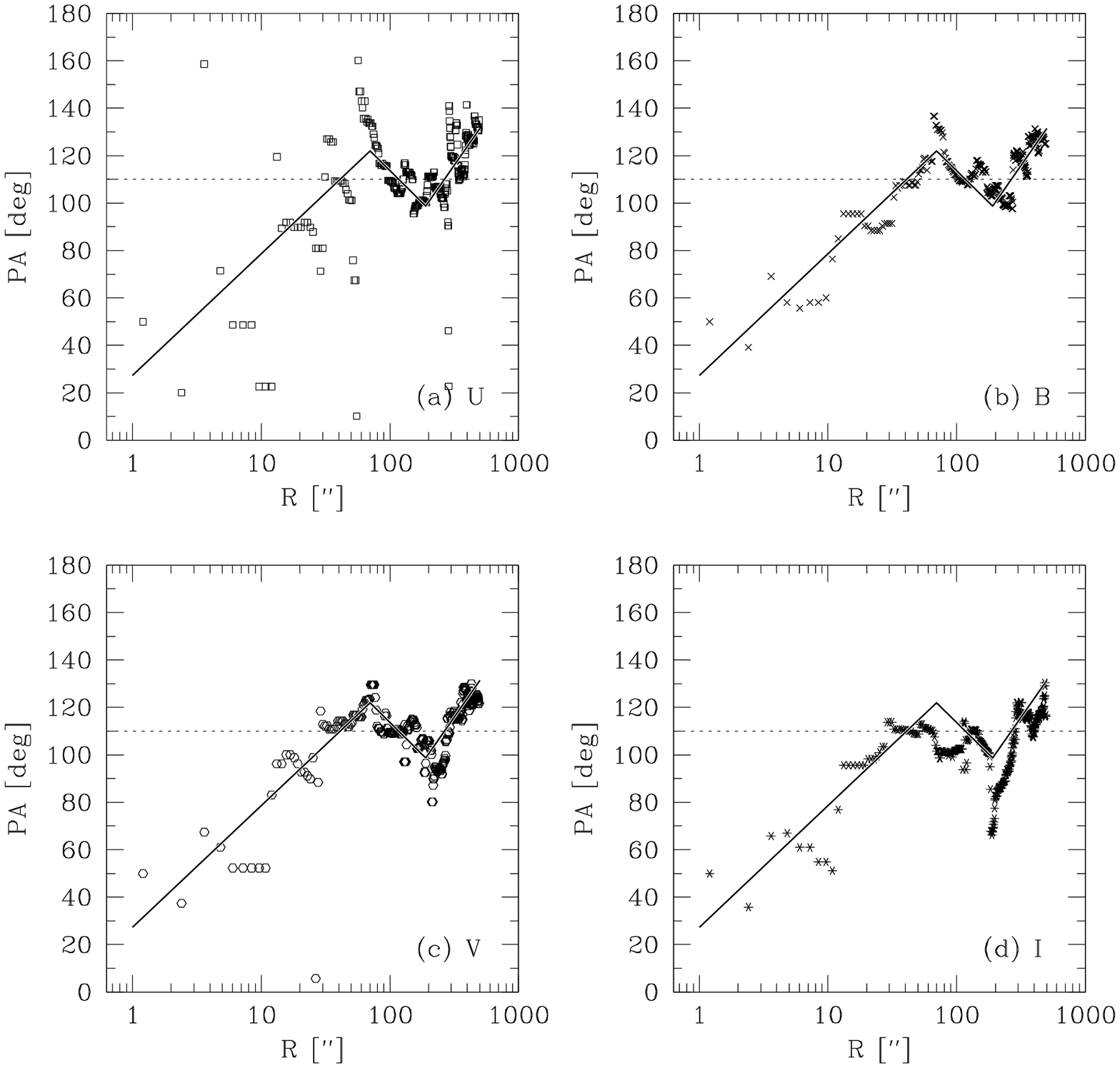}
   \caption{Position angle profiles of NGC 300.
The straight lines are based on eye-fitting and are the same for
all bands.}
   \end{center}
\end{figure}

\section{Color Profiles}  
Fig. 6 shows the radial surface color profiles of NGC 300
along the major axis. 
Here the surface color means the differential color per square arcsecond.
The errors for $(B-V)$ color are ploted at the upper panel
in order to show the typical error values.
The solid lines are least-square fits to the data that yield the color gradient slopes of 
$-0.07$ mag arcsec$^{-2}$ (100\arcsec)$^{-1}$ for $(U-B)$, 
$-0.11$ mag arcsec$^{-2}$ (100\arcsec)$^{-1}$ for $(U-V)$, 
$-0.14$ mag arcsec$^{-2}$ (100\arcsec)$^{-1}$ for $(U-I)$,
$-0.04$ mag arcsec$^{-2}$ (100\arcsec)$^{-1}$ for $(B-V)$,
$-0.07$ mag arcsec$^{-2}$ (100\arcsec)$^{-1}$ for $(B-I)$, and
$-0.04$ mag arcsec$^{-2}$ (100\arcsec)$^{-1}$ for $(V-I)$.
As the wavelength separation between the two filters gets larger,
the slope of the corresponding color profile gets steeper.
Here, the steepest slope is for the $(U-I)$ color, and
the least steep slopes are for the $(B-V)$ and $(V-I)$ colors.

   \begin{figure}
   \begin{center}
   \plotone{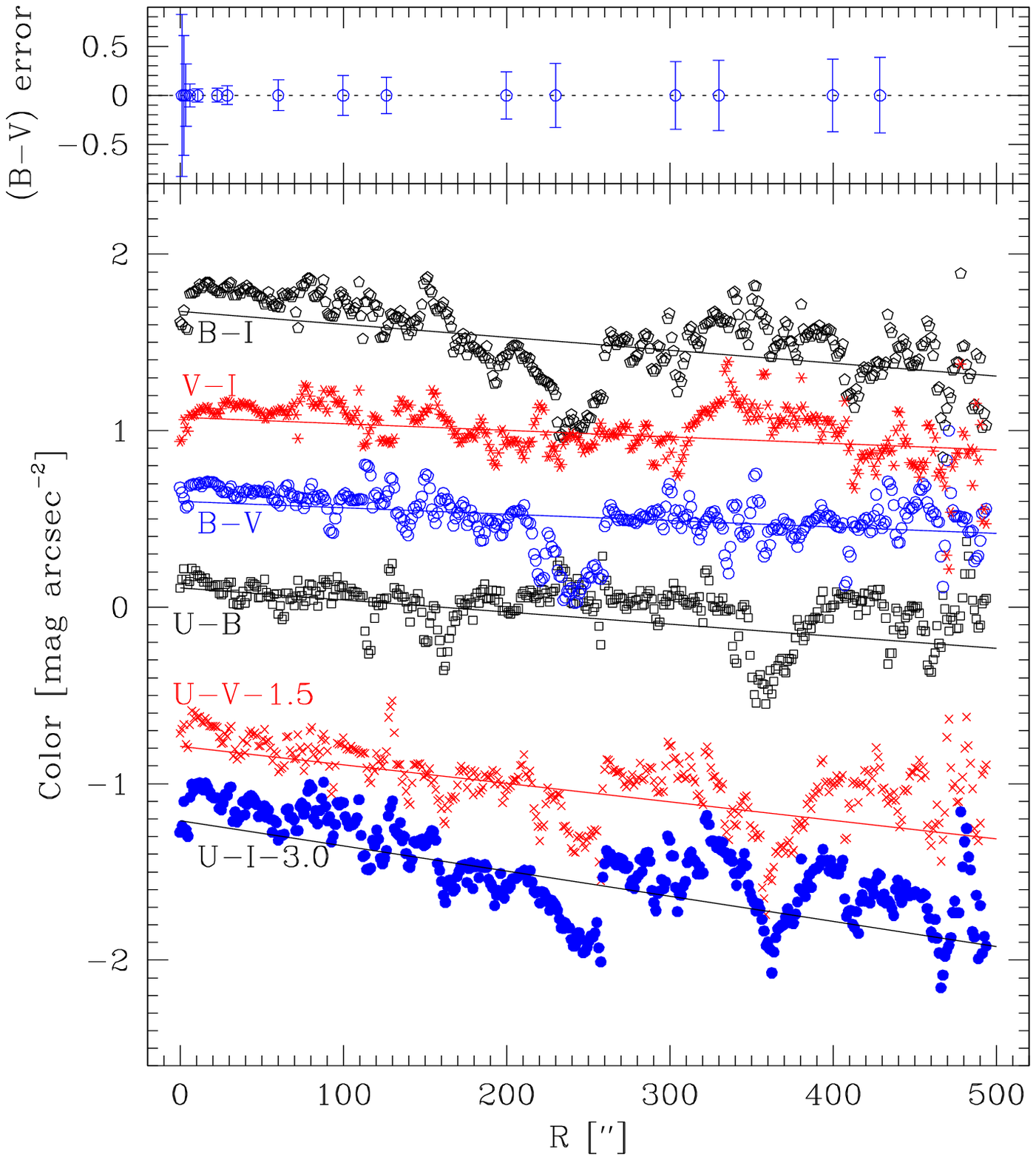}
   \caption{Radial surface color profiles of NGC 300.
The errors for $(B-V)$ color are ploted at the upper panel
to show the typical error values.}
   \end{center}
\end{figure}

NGC 300 shows typical color profiles of late-type spiral galaxies,
being redder in the inner regions and getting bluer as the radius increases.
The primary factor for this negative radial color gradient is 
the differences of the underlying stellar populations at the inner and outer
parts of the galaxies (Kim \& Ann 1990).

\section{Discussion} 

\subsection{Comparison with Previous Studies} 
  Fig. 7 shows the comparison of the surface brightness
profiles obtained in this study with those of previous studies.
Fig. 7 (a) is in logarithmic radius scale and
Fig. 7 (b) is in linear radius scale.
$B$- and $I$-band surface brightness profiles obtained in this study are denoted
as open squares and open circles, respectively.
Both de Vaucouleurs \& Page (1962) and Carignan (1985) have given $B$-band 
surface brightness profiles obtained from photographic plates,
which are denoted as dashed lines and dotted lines (equivalent profile) or
open triangles (elliptically averaged profile), respectively.
$B_J$ magnitudes of Carignan (1985) are transformed to $B$ magnitudes using
the relation $B = B_J + 0.06(B_J - R_F) = B_J +0.05$ (Carignan 1985).
Recent {\it HST}-based $I$-band surface brightness profiles for the inner
$r < 15$\arcsec~ region of NGC 300 given by B\"oker et al. (2002) are denoted 
by solid lines.  
Because the B\"oker et al. (2002)'s $I$-band surface brightnesses are 
0.58 mag arcsec$^{-2}$ brighter than those in this study, we have added
0.58 mag arcsec$^{-2}$ to their values to match the two profiles.

\begin{figure}
   \begin{center}
   \plottwo{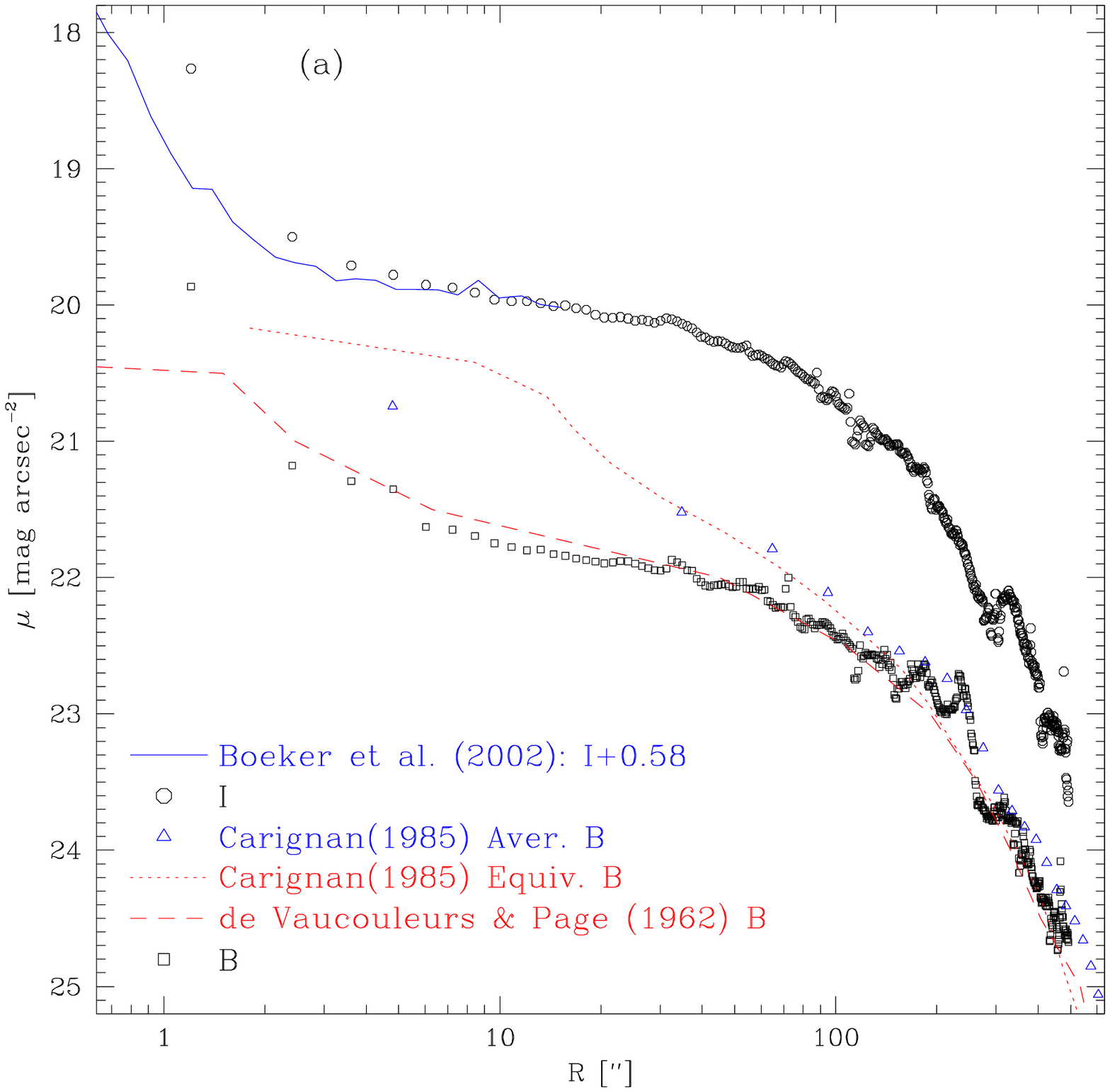}{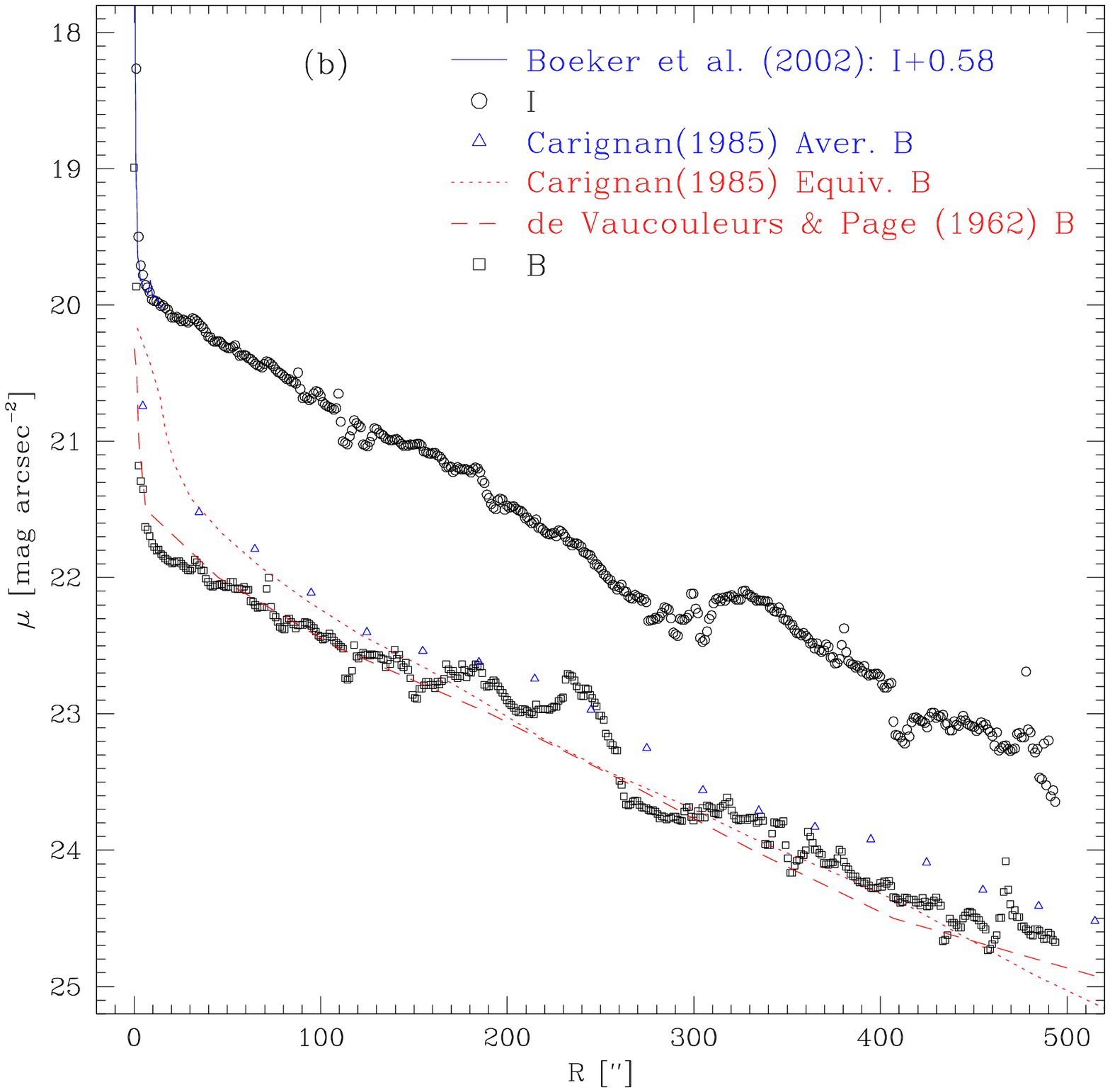}
   \caption{Comparison of the surface brightness profiles
obtained in this study with those of previous studies in
logarithmic radius scale (panel a) and
linear radius scale (panel b).  $B$ and $I$ surface brightnesses obtained
in this study are denoted as open squares and open circles, respectively.
$B_J$ magnitudes of Carignan (1985) are transformed to $B$ magnitudes using
the relation $B = B_J + 0.06(B_J - R_F) = B_J +0.05$ (Carignan 1985).
0.58 mag arcsec$^{-2}$ is added to the B\"oker et al. (2002)'s $I$-band
surface brightnesses to match their profiles to those obtained in
this study.}
   \end{center}
\end{figure}

  For the $B$-band surface brightness profiles, our profiles are in good 
agreement with those of de Vaucouleurs \& Page (1962) at $r$ \simlt 160\arcsec~
and with equivalent profiles of Carignan (1985) at the outer region.
Carignan (1985) notes the differences between his and de Vaucouleurs \& Page (1962)'s
surface brightness profiles, which is up to 0.4 mag arcsec$^{-2}$ and 
is larger at the central region.
This difference between the studies of de Vaucouleurs \& Page (1962) 
and Carignan (1985) seems to be due to some problems in 
the density-to-intensity calibration of the photographic plates
since both studies are based on photographic plate photometry.

  The $I$-band surface brightness profiles of NGC 300 given by B\"oker et al. 
(2002) are $\sim$0.6 mag arcsec$^{-2}$ brighter than those of our study.
Since B\"oker et al. (2002) have observed their galaxies with the F814W
filter only, they have made the photometric calibration and conversion 
to Johnson $I$-band assuming a standard color of $V-I = 1$ for all the galaxies.
Nevertheless, this does not account for the rather large difference 
between the two photometric studies, because even 1 mag of change in the color of 
a galaxy gives only $\sim$ 0.01 mag difference of zero point as they mentioned.
Apart from the zeropoint difference between the two studies and 
some discrepancies of the values at a few central points,
the two $I$-band surface brightnesses show reasonable agreement especially 
at $r > 5$\arcsec.
For the very central regions of $r < 5$\arcsec, B\"oker et al. (2002)'s results
based on the {\it HST} high-resolution CCD images deserve to be considered 
as reliable.

\subsection{Surface Brightness Profile Decomposition} 

  We have combined our $I$-band surface brightness measurements for the region
of $r > 5$\arcsec~ with those
of B\"oker et al. (2002) for the region from 0.\arcsec02 to 5\arcsec.
The resulting $I$-band surface brightness profile is shown in Fig. 8
as open circles.
  Using this combined surface brightness profile for 
the region of 0.\arcsec02 $< r <$ 500\arcsec,
we have performed a three component decomposition: 
a nucleus, a bulge and a disk (Stephens \& Frogel 2002).
  We assume an exponential disk 
\begin{equation}
I_{disk}(r) = I_0~ {\rm exp}\frac{-r}{R_d}
\end{equation}
with a central intensity $I_0$ and a disk scale length $R_d$, and 
a de Vaucouleurs $r^{1/4}$ bulge 
\begin{equation}
I_{bulge}(r) = I_e~ {\rm exp}\{ -7.67~ [(\frac{ar}{R_e})^{1/4}-1]\}
\end{equation}
with an effective radius $R_e$ and an intensity $I_e$ at $R_e$.
  The factor $a = [(1+{\rm cos}^2~ i)/2]^{1/2}$ is a correction factor for
deprojecting a spherical bulge, and an inclination of $i = 42.\arcdeg3$ 
(Carignan 1985) is used.
  For the core component, we have used Sersic model (Sersic 1968):
\begin{equation}
I_{core}(r) = I_s~ {\rm exp}\{ -b_n [(\frac{ar}{R_s})^{1/n} -1]\},
\end{equation}
where $R_s$ is an effective radius, $I_s$ is an intensity at $R_s$, and
$n$ is the shape index ($b_n \approx 1.9992n - 0.3271$).
The result of the three-component decomposition is shown in Fig. 8,
and the parameter values are listed in Table 2.
Fig. 8 shows that NGC 300 has three components: the 
central peak in surface brightness at $r < 1$\arcsec~ which is reasonably fitted
by the Sersic function, 
the disk that has a central surface brightness of 19.97 mag arcsec$^{-2}$
and scale length of $R_d =$ 2.\arcmin5 (= 1.47 kpc), and
a very faint spheroidal component which is even fainter than
that of M33 (Stephens \& Frogel 2002).

  The nuclear star cluster observed by B\"oker et al. (2002) 
using the {\it HST} has been included in the globular cluster candidate 
catalogue in Paper I with highest probability.
It would be helpful to perform a spectroscopic observation on this 
nuclear star cluster of NGC 300 to confirm the suggestion of Soffner et al. 
(1996) that the nucleus of this galaxy is an unresolved compact stellar
cluster and to understand the nature of it.
Information on the velocity dispersion of this object will also give
a clue on the existence of any black hole in it (Gebhardt et al. 2001;
Gebhardt, Rich, \& Ho 2002).

More surface brightness data in other passbands at high spatial resolution
will be valuable to get 
information on the existence of any color gradient in the nuclear region 
of NGC 300 (Kormendy \& McClure 1993; Kim \& Lee 1998).

\begin{figure}
   \begin{center}
   \plotone{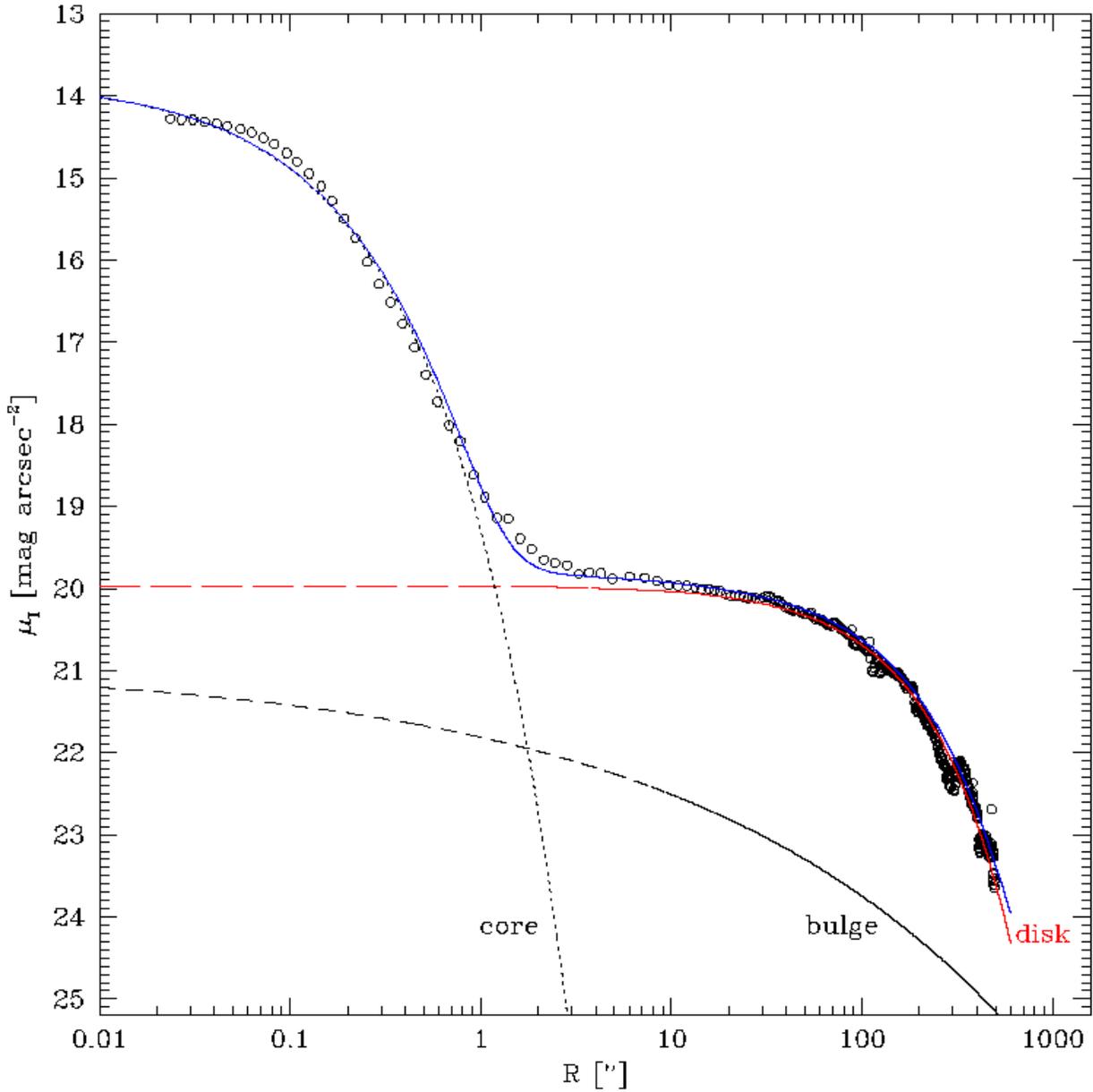}
   \caption{Profile decomposition of $I$-band surface brightness profile
obtained in this study (for $r > 5$\arcsec)
combined with measurements from B\"oker et al. (2002; 0.\arcsec02 $< r < $5\arcsec).
The open circles are the observed data points,
and the results of a three-component exponential disk ({\it long-dashed line}),
$r^{1/4}$ bulge ({\it short-dashed line}) and Sersic core ({\it dotted
line}) decomposition.
The solid line is the combined result of the three model components.}
   \end{center}
\end{figure}

\begin{table*}
\begin{center}
{\footnotesize
\caption{Model Parameters for $I$-band Three-Component Decomposition}
\begin{tabular}{l|lc}
  \hline
  & Parameter & Value \\
  \hline
Disk & scale length, $R_d$                  & 150\arcsec $=$ 2.\arcmin50 = 1.47 kpc$^{\rm a}$ \\
     & central surface brightness, $\mu_0$  & 19.97 mag arcsec$^{-2}$ \\
  \hline
Bulge& effective radius, $R_e$              & 1900\arcsec $=$ 31.\arcmin67 = 18.62 kpc$^{\rm a}$ \\
     & surface brightness at $R_e$, $\mu_e$ & 29.25 mag arcsec$^{-2}$ \\
  \hline
Core & effective radius, $R_s$              & 0.\arcsec094 = 0.92 pc$^{\rm a}$ \\
     & surface brightness at $R_s$, $\mu_s$ & 16.55 mag arcsec$^{-2}$ \\
     & shape index, $n$                     & 1.43 \\
\hline
\end{tabular}
} 
\\
Note. $^{\rm a}$ Assuming the distance modulus of (m-M)$_0 = 26.53 \pm
0.07$ mag (Freedman et al. 2001), which gives 1\arcsec = 9.8 pc.
\end{center}
\end{table*}

\section{Summary} 

We have presented surface photometry for the $20.'5 \times 20.'5$ area of 
SA(s)d galaxy NGC 300 in the Sculptor group.
We have derived the isophotal maps, surface brightness profiles, 
ellipticity profiles, position angle profiles, and color profiles.
We have decomposed the $I$-band surface brightness profiles of
0.\arcsec02 $< r <$ 500\arcsec~ region into a nucleus, a bulge, and
an exponential disk.

\section*{Acknowledgments}
We thank the anonymous referee for the constructive comments and suggestions
that have improved this paper.
   S.C.K. is grateful
to Mr. Yoon-Ho Park and Mr. Chang Hyun Baek for providing information on 
the {\small SPIRAL} package,
and to Dr. Bong Gyu Kim for his continuous and graceful encouragements 
on astronomical research.
   This research has made use of the NASA/IPAC Extragalactic Database (NED) which 
is operated by the Jet Propulsion Laboratory, California Institute of Technology, 
under contract with the National Aeronautics and Space Administration. 
  H. S. acknowledges the support of the Korea Science and Engineering Foundation 
to the Astrophysical Research Center for the Structure and Evolution of the Cosmos 
(ARCSEC) at Sejong University.
The data used in this paper are available by e-mail request to {\it sckim@kao.re.kr}.



\begin{thebibliography}{}  
\bibitem{} Blair W. P., Long K. S., 1997, ApJS, 108, 261
\bibitem{} B\"oker T., Laine S., van der Marel R. P., Sarzi M., Rix H.-W.,
        Ho L. C., Shields J. C., 2002, AJ, 123, 1389
\bibitem{} B\"oker T., Sarzi M., McLaughlin D. E., van der Marel R. P., Rix H.-W.,
        Ho L. C., Shields J. C., 2004, AJ, 127, 105
\bibitem{} Butler D. J., Mart{\'i}nez-Delgado D., Brander W., 2004, AJ, 127, 1472
\bibitem{} Carignan C., 1985, ApJS, 58, 107
\bibitem{} Choi Y.-J., Park B.-G., Yoon T. S., Ann H. B., 1998, Jour. of 
        the Korean Astron. Soc., 31, 141
\bibitem{} C{\^o}t{\'e} S., Freeman K. C., Carignan C., Quinn P. J.,
        1997, AJ, 114, 1313  
\bibitem{} de Vaucouleurs G., Page J., 1962, ApJ, 136, 107
\bibitem{}
        de Vaucouleurs G., de Vaucouleurs A., Corwin H. G., Jr.,
        Buta R. J., Paturel G., Fouqu{\'e} P., 1991,
       Third Reference Catalogue of Bright Galaxies,
       New York: Springer-Verlag (RC3)
\bibitem{}
        Elias J., 1994, NOAO Newsletter, No. 37, 1
\bibitem{}
        Freedman W. L., et al., 2001, ApJ, 553, 47
\bibitem{} Gebhardt K., et al., 2001, AJ, 122, 2469
\bibitem{} Gebhardt K., Rich R. M., Ho L. C., 2002, ApJ, 578, L41
\bibitem{} Ichikawa S., Okamura S., Watanabe M., Hamabe M., Aoki T., 
        Kokaira K., 1987, Annals Tokyo Astron. Obs., 21, 285
\bibitem{} Karachentsev I. D., Grebel E. K., Sharina M. E., Dolphin A. E., 
        Geisler D., Guhathakurta P., Hodge P. W., Karachentseva V. E., 
        Sarajedini A., Seitzer P., 2003, A\&A, 404, 93
\bibitem{} Kent S. M., 1983, ApJ, 266, 562
\bibitem{}
        Kilkenny D., van Wyk F.,
        Roberts G., Marang F., Coper D., 1998, MNRAS, 294, 93
\bibitem{}
        Kim E., Lee M. G., \& Geisler D., 2000, MNRAS, 314, 307
\bibitem{} Kim K. O., Ann H. B., 1990, Jour. of the Korean Astron. Soc., 
        22, 43
\bibitem{} Kim S. C., Lee M. G., 1998, Jour. of the Korean Astron. Soc., 31, 51
\bibitem{}
        Kim S. C., Sung H., Lee M. G., 2002, Jour. of the Korean
        Astron. Soc., 35, 9 (Paper I)
\bibitem{} Kormendy J., McClure R. D., 1993, AJ, 105, 1793
\bibitem{} Leinert Ch., V\"ais\"anen P., Mattila K., \& Lehtinen K.,
        1995, A\&AS, 112, 99
\bibitem{}
        Menzies J., Marang F., Laing J. D., Coulson I. M., 
        Engelbrecht C. A., 1991, MNRAS, 248, 642
\bibitem{}
        Pietrzy\'nski G., Gieren W., Fouqu\'e P., Pont F., 2001, 
        A\&A, 371, 497
\bibitem{} Read A. M., Ponman T. J., Strickland D. K., 1997,
        MNRAS, 286, 626
\bibitem{} Richer H. B., Pritchet C. J., Crabtree D. R., 1985,
        ApJ, 298, 240
\bibitem{} Rogstad D. H., Crutcher R. M., Chu K., 1979, ApJ, 229, 509
\bibitem{}
        Sandage A., Bedke J., 1994, The Carnegie Atlas of Galaxies, 
        Washington, D. C.: Carnegie Institution of Washington
\bibitem{} Sandage A. R., Tammann G. A., 1981, A Revised Shapley-Ames
        Catalog of Bright Galaxies, Washington, D. C.: Carnegie Institution
        of Washington
\bibitem{}
        Schlegel D., Finkbeiner D., Davis M., 1998,
        ApJ, 500, 525
\bibitem{} Sersic J. L., 1968, Atlas de Galaxias Australes, C\'ordoba: Observatorio
        Astron\'omico
\bibitem{} Soffner T., M\'endez R. H., Jacoby G. H., Ciardullo R.,
        Roth M. M., Kudritzki R. P., 1996, A\&A, 306, 9
\bibitem{} Stephens A. W., Frogel J. A., 2002, AJ, 124, 2023
\bibitem{}
        Sung H., Bessell M. S., 2000, PASA, 17, 244
\bibitem{} van den Bergh S., 1999, ApJ, 517, L97
\bibitem{} Whiting A. B., 1999, AJ, 117, 202
\end{thebibliography}
\end{document}